\documentclass[3p,12pt]{elsarticle}
\bibstyle{plane}
\biboptions{sort&compress}
\usepackage{amssymb}
\usepackage{color,slashed,mathrsfs}
\usepackage{natbib}

\usepackage{amssymb,color,graphicx,bm}
\newcommand{\ie}{{\it i.e., }}

\newcommand{\vecB}{{\bm B}}
\newcommand{\vecM}{{\bm M}}

\definecolor{red}{rgb}{0.8,0,0}
\definecolor{RED}{rgb}{0.8,0,0}
\definecolor{violet}{rgb}{0.4,0,0.4}
\definecolor{green}{rgb}{0,0.5,0.0}
\definecolor{GREEN}{rgb}{0,0.5,0.0}
\definecolor{navy}{rgb}{0.0,0.0,0.6}
\definecolor{orange}{rgb}{0.8,0.2,0.0}
\definecolor{blue}{rgb}{0.3,0.0,0.8}

\begin{document}

\begin{frontmatter}
\title{Hypernuclear matter  in strong magnetic field}  
\author{Monika Sinha}
\address{
 Institute for Theoretical Physics, J. W.\ Goethe-University, D-60438 Frankfurt am\
 Main, Germany
}
\address{
 Indian Institute of Technology Rajasthan, Old Residency Road, 
Ratanada, Jodhpur 342011, India
}
\author{Banibrata Mukhopadhyay}
\address{Department of Physics, Indian Institute of Science, 
       Bangalore 560012, India}
\author{Armen Sedrakian}
\address{Institute for Theoretical Physics,
J.~W.~Goethe University, D-60438  Frankfurt-Main, Germany}

\date{today}

\begin{abstract}
  Compact stars with strong magnetic fields (magnetars) have been
  observationally determined to have surface magnetic fields of order of
  $10^{14}-10^{15}$ G, the implied internal field strength being
  several orders larger. We study the equation of state and
  composition of dense hypernuclear matter in strong magnetic fields
  in a range expected in the interiors of magnetars. Within the
  non-linear Boguta-Bodmer-Walecka model we find that the magnetic
  field has sizable influence on the properties of matter for central
  magnetic field $B \ge 10^{17}$~G, in particular the matter
  properties become anisotropic. Moreover, for the central fields
  $B \ge 10^{18}$~G, the magnetized hypernuclear matter
  shows instability, which is signalled by the negative sign of the
  derivative of the pressure parallel to the field with respect to the density,
  and leads to vanishing parallel pressure at the critical value
  $B_{\rm cr} \simeq  10^{19}$~G.
   This limits the range of admissible homogeneously distributed fields in
  magnetars to fields below the critical value $B_{\rm cr}$.
\end{abstract}

\end{frontmatter}

\section{Introduction}

Anomalous X-ray pulsars and soft $\gamma$-ray repeaters are
observationally identified with highly magnetized neutron stars 
with surface magnetic field $\sim 10^{14}-10^{15}$ G
\cite{1997ApJ...486L.129V,1998Natur.393..235K,1999ApJ...519L.139W}. This class of compact stars has been
conjectured theoretically as ``magnetars''~\cite{1992ApJ...392L...9D,1992Natur.357..472U,1995MNRAS.275..255T,1996ApJ...473..322T},
\ie neutron stars which posses magnetic fields that are by several
orders of magnitude larger than the canonical surface dipole magnetic
fields $B\sim 10^{12}$-$10^{13}$ G of the bulk of pulsar population,
which are commonly deduced from the magnetic-dipole radiation model of
pulsar spin-down in combination with measured spin and spin-down
rates.  The integral properties of magnetars, \ie mass, radius, moment
of inertia, {\it etc} will depend sensitively on the equation of state
of matter in strong magnetic fields, if the central fields are
sufficiently strong.  Furthermore, other processes, such as the cooling
and the magnetic field evolution will sensitively depend on the
composition of matter in strong magnetic fields. Fermionic matter in
strong magnetic field experiences two well-known quantum mechanical
phenomena: the Pauli paramagnetism and the Landau diamagnetism. The
first is due to the interaction of the spin of the fermion with the
magnetic field and, therefore, is relevant for both charged and
uncharged fermions. The second phenomenon is relevant only for charged
particles, and is particularly strong for light particles, which in the
case of compact stars are the leptons.

Neutron star sequences may feature massive objects with masses $M\sim
2M_{\odot}$.  These massive compact stars are likely to develop cores
which are composed of matter that differs from the ordinary nuclear
matter composed of only neutrons and protons. One possibility is that
heavy baryons (hyperons) will appear once the Fermi energy of neutrons
becomes of the order of their rest mass. Although the hyperons were
considered even before the discovery of pulsars and their
identification with the neutron stars~\cite{1960AZh....37..193A},
their presence in the cores of neutron stars is still quite uncertain,
for different theoretical models predict quite different outcomes for
the maximum mass of hypernuclear compact stars, often in contradiction
with the known empirical data on pulsar masses.  The models that are
based on relativistic density functional
methods~\cite{1985ApJ...293..470G,1991PhRvL..67.2414G,1999paln.conf.....W,2007PrPNP..58..168S}
predict masses that are not much larger than the canonical mass of a
neutron star which clearly contradicts modern observations. Masses of
the order of $\le 1.8 M_{\odot}$ were obtained in non-relativistic
phenomenological
models~\cite{1999ApJS..121..515B,2010PhRvC..81c5803D}, while
microscopic models based on hyperon-nucleon potentials, which include
the repulsive three-body forces, predict low maximal masses for
hypernuclear stars~\cite{2003astro.ph.12446B,2011EL.....9411002V}.
The avenues for reconciliation of the large pulsar mass and the
hyperonization (and more generally strangeness) of dense matter have 
been explored recently in 
Refs~\cite{2012A&A...539A..16B,2012ApJ...756...56J,2012A&A...543A.157B,2012PhRvC..85e5806O,2012arXiv1205.3621M,2011arXiv1112.6430L,2011arXiv1112.1853S,2012arXiv1206.3086D,2012NuPhA.881...62W,2012PhRvC..85f5802W,2012arXiv1209.0270G,2012arXiv1207.4872M}.

The influence of strong magnetic fields on the highly dense electron
gas in the context of neutron star matter was studied in Refs.
\cite{1977FCPh....2..203C,1989ApJ...342..958F,1991ApJ...374..652A,1992AnPhy.216...29F,1993ApJ...416..276R}.
The strong magnetic field effects on dense nuclear matter ($n, p, e$
system) have been studied previously in
Refs. \cite{1997PhRvL..78.2898C,1998PhRvD..58l1301B,2000ApJ...537..351B,2001ApJ...546.1126S,2006RPPh...69.2631H,2007MPLA...22..623C,2008JPhG...35l5201R}. The
structure of strongly magnetized neutron star branch of compact
objects was studied for different field configurations (toroidal,
poloidal, etc) in
Refs.~\cite{1995A&A...301..757B,2001ApJ...554..322C,2012arXiv1207.4035F}.
In the case of the adjacent branch of white dwarfs, strong magnetic
fields were found to lead to highly super-Chandrasekhar mass ($M\sim
2.3-2.6M_\odot$) white
dwarfs~\cite{2012PhRvD..86d2001D,2012IJMPD..2142001D,2012MPLA...2750084K},
which can be related to the observed features of a number of peculiar
Type Ia supernovae~\cite{2010ApJ...713.1073S}.

It has been known for some time that the magnetic field can affect the
hydrostatic equilibrium of compact stars and may render large fields
configurations unstable.  In the simplest form it can be formulated
for uniform self-gravitating fluids~\cite{1953ApJ...118..116C}.  In
the case of neutron stars the Chandrasekhar-Fermi limiting field
strength is $\sim 10^{18}$~G~\cite{1991ApJ...383..745L}. Fully
relativistic calculations confirm the simple Newtonian estimates
~\cite{1995A&A...301..757B,2001ApJ...554..322C,2012arXiv1207.4035F}.
Instabilities related to the anisotropy of the pressure were also
discussed in the literature. These types of instabilities may 
arise due to the change of the sign of the derivative of either 
the transverse pressure or the parallel pressure with respect 
to the density, which leads eventually to vanishing of the 
respective pressure. These instabilities have been discussed 
for electron gas~\cite{2000PhRvL..84.5261C} and strange quark 
matter~\cite{2008IJMPD..17.2107P,2010PhRvD..81d5015H,2011AnPhy.326.3075H} 
due to the vanishing of transverse pressure and for magnetized 
fermionic systems~\cite{2010PhRvC..82f5802F} and quark matter
~\cite{2011PhRvD..83d3009P,2011arXiv1108.4479D} due to parallel 
pressure (but see \cite{PhysRevC.85.039801,2012PhRvC..85c9802F}).

The studies of magnetized dense matter were mainly carried out in the
limit of uniform field distribution, some notable exceptions are
Refs.~\cite{1997PhRvL..79.2176B,1998PhRvD..58l1301B,2010PhRvC..82f5802F}.
The processes of supernova collapse will leave behind a strongly
non-uniform field distribution of frozen-in fields.  Any dynamo
mechanism generating fields will carry the imprint of inhomogeneous
density profile in the star. Thus, more realistic treatment of the
equation of state of matter requires inclusion of some physically
motivated field profiles of the stars such as magnetars. (We note that
{\it local} equation of state is not affected by the density profile
of the field, and therefore depends only on the local value of the
magnetic field strength and induction).  Thus, while uniform magnetic
field distribution gives a crude order of magnitude estimates, it
requires refinements, because the density profiles of neutron stars
are not uniform. Consequently, the local magneto-static equilibrium
will be achieved for different field strength.

The motivation of this paper is to carry out a study which includes
the following aspects of the physics of strong magnetic field
discussed above: (i) we allow for radial profile of the magnetic
field, which assumes a high-field central value which decays as the
field stretches towards the surface. In this way, {\it we extend the
  existing studies of hypernuclear matter to the case of non-uniform
  fields}. (ii) We carefully analyze the different components of the
field and show, in our case study of hypernuclear matter, that {\it
  the parallel pressure shows the instability}.
Thus, our work combines
and unifies the ideas of the non-homogenous distribution of magnetic
fields in magnetars and possible instabilities in a new context of
nuclear and hypernuclear matter. We quantify these effects by studying
the dependence of the equation of state, onset of instability, and
composition of matter by varying our parameterization of the field
profiles. 

This paper is organized as follows. In Sec.~\ref{model} we discuss our
model for hypernuclear matter in strong magnetic fields within a
relativistic density functional theory. We present our numerical
results in Sec.~\ref{discussion}.  Section~\ref{summary} contains a
summary of our results.

\section{Model}\label{model}

As discussed in the introduction, the treatment of hypernuclear matter
within nuclear models which are based on different principles has lead
to array of results, some of which are inconsistent with measurements
of pulsar masses. Relativistic models of hypernuclear matter are
flexible enough to be adjusted to the current phenomenology, therefore
they provide a suitable framework for extensions which incorporate
the influence of strong magnetic fields.  In particular, we consider
the non-linear Boguta-Bodmer-Walecka model which is based on
relativistic density functional theory~
\cite{1974AnPhy..83..491W,1977NuPhA.292..413B,1982PhLB..114..392G,1985ApJ...293..470G,1987ZPhyA.326...57G,1987ZPhyA.327..295G}. 
This model is able to describe ground state phenomenology of 
nuclear matter and nuclei (see, e.g., \cite{1977AnPhy.108..301C,Serot:1984ey}). 
We anticipate that the main conclusions and results of the 
present study will not change qualitatively if a different 
model of dense hypernuclear matter is chosen.

The Lagrangian density of hypernuclear matter in a static magnetic
field (hereafter also $B$-field) is given by 
\begin{equation}
\label{eq:Ldens}
{\cal L} =  {\cal L}_m +  {\cal L}_f  \label{lag},
\end{equation}
where the matter part of the Lagrangian is given by 
\begin{eqnarray}
{\cal L}_m &=& \sum_B \bar\psi_{B}\left(i\gamma_\mu 
D^{\mu} - m_B
+ g_{\sigma B} \sigma - g_{\omega B} \gamma_\mu \omega^\mu
- g_{\rho B}
\gamma_\mu{\mbox{\boldmath $\tau$}}_B \cdot
{\mbox{\boldmath $\rho$}}^\mu \right)\psi_B\nonumber\\
&& + \frac{1}{2} \partial_\mu \sigma\partial^\mu \sigma
- \frac12 m_\sigma^2 \sigma^2 - U(\sigma) \nonumber\\
&& -\frac{1}{4} \omega_{\mu\nu}\omega^{\mu\nu}
+\frac{1}{2}m_\omega^2 \omega_\mu \omega^\mu
- \frac{1}{4}{\mbox {\boldmath $\rho$}}_{\mu\nu} \cdot
{\mbox {\boldmath $\rho$}}^{\mu\nu}
+ \frac{1}{2}m_\rho^2 {\mbox {\boldmath $\rho$}}_\mu \cdot
{\mbox {\boldmath $\rho$}}^\mu \nonumber \\
&& + \sum_{l=e,\mu} \bar\psi_{l}\left(i\gamma_\mu 
D^{\mu} - m_l \right)\psi_l, 
\label{lag}
\end{eqnarray}
where the index $B$ labels baryons present in the matter,
$\psi_B,\psi_l,\sigma,\omega$ and ${\mbox {\boldmath$\rho$}}$ are
fields of baryons, leptons, $\sigma$-mesons, $\omega$-mesons and
$\rho$-mesons, with masses $m_B, m_l, m_\sigma, m_\omega$ and $m_\rho$
respectively; $g_{\sigma B}, g_{\omega B}$ and $g_{\rho B}$ are
coupling constants for interactions of $\sigma, \omega$ and $\rho$
mesons respectively with the baryon $B$.
The scalar self interaction term in the matter Lagrangian is~\cite{1977NuPhA.292..413B,1987ZPhyA.327..295G,1987ZPhyA.326...57G,1982PhLB..114..392G,1985ApJ...293..470G} 
\begin{equation}
U(\sigma)~=~\frac13~g_1~m_N~(g_{\sigma N}\sigma)^3~+~ \frac14~g_2~
(g_{\sigma N}\sigma)^4~,
\end{equation}
where $g_1$ and $g_2$ are constants which parameterize the shape of
the potential and
\begin{equation}
\omega_{\mu\nu}~=~\partial_\nu\omega_\mu~-~\partial_\mu\omega_\nu,
\end{equation}
\begin{equation}
{\mbox{\boldmath $\rho$}}_{\mu\nu}~=~\partial_\nu{\mbox{\boldmath $\rho$}}_\mu~-
~\partial_\mu{\mbox{\boldmath $\rho$}}_\nu,
\end{equation}
\begin{equation}
D^{\mu}=\partial^{\mu}+ieQA^{\mu}.
\end{equation}
We choose the field vector potential gauge as
$A^{\mu}\equiv(0,-yB,0,0)$, where $B$ is the magnitude of magnetic
field and $eQ$ the charge of the particle, $e$ the (positive) unit of
charge. For this particular gauge choice the magnetic field is along
the $z$ axis of Cartesian coordinate system, \ie
${\vecB}={B}\hat{z}$. In the mean field approximation, the baryons
acquire the effective masses 
\begin{equation}
m_B^*~=~m_B~-~g_{\sigma B} \sigma,
\end{equation}
where $\sigma$ is given by its ground state expectation value
\begin{equation}
<\sigma>=\sigma~=~\frac1{m_\sigma^2}\left(\sum_B g_{\sigma B} ~ n_S^{(B)}
~-~\frac{\partial U}{\partial \sigma}\right),
\end{equation}
with the scalar density
\begin{equation}
n_S^{(B)}~=~\frac2{(2\pi)^3} \int_0^{p_F^{(B)}} 
\frac{m_B^*}{\sqrt{p_B^2+m_B^{*^2}}} d^3p_B,
\end{equation}
where $p_B$ is the momentum and $p_F^{(B)}$ the Fermi momentum of the
baryon $B$.
The electromagnetic field Lagrangian density is given by
\begin{equation}
{\cal L}_f  =  -\frac1{16\pi} F_{\mu \nu}F^{\mu \nu}, \label{lagf}
\end{equation}
where $F^{\mu\nu}$ is the  electromagnetic field 
tensor.

The total energy density and pressure of the system can be obtained
by computing the energy-momentum tensor from the Lagrangian density (\ref{eq:Ldens}).
The result is 
\begin{equation}
T^{\mu \nu} = T^{\mu \nu}_m + T^{\mu \nu}_f,
\end{equation}
where the matter part is given by 
\begin{equation}
T^{\mu\nu}_m= \varepsilon_m u^\mu u^\nu - P(g^{\mu\nu}-u^\mu u^\nu)
+ \frac12 (M^{\mu \lambda}F_\lambda^\nu + M^{\nu \lambda}F_\lambda^\mu),
\label{tmatter}
\end{equation}
with $\varepsilon_m$ being the matter energy density, $P$ - the
thermodynamic pressure, $M^{\mu \nu}$ - the magnetization tensor. The
field part of the energy-momentum tensor is given by 
\begin{equation}
T^{\mu\nu}_f = -\frac{1}{4\pi}F^{\mu\lambda}F^\nu_\lambda + \frac1{16\pi} g^{\mu \nu} 
F^{\rho\sigma} F_{\rho\sigma}. 
\label{tmag}
\end{equation}
In the following we will neglect the electric field, as there are no
macroscopic charges in the bulk matter.  Therefore,
Eqs. (\ref{tmatter}) and (\ref{tmag}) reduce, respectively, to
\cite{2002PhRvD..65e6001K,2010PhRvD..81d5015H}
\begin{eqnarray}
T^{\mu\nu}_m &=& \varepsilon_m u^\mu u^\nu - P(g^{\mu\nu}-u^\mu u^\nu)
+ M {B} \left(g^{\mu\nu}-u^\mu u^\nu + \frac{{B}^\mu {B}^\nu}{{B}^2}\right),\\
T^{\mu \nu}_f &=& \frac{{B}^2}{4\pi} \left(u^\mu u^\nu - \frac12 g^{\mu\nu}\right) 
- \frac{{B}^\mu {B}^\nu}{4\pi},
\end{eqnarray}
with ${B}^\mu{B}_\mu = - {B}^2$ and $M$ being the magnetization 
per unit volume. 

In the presence of the magnetic field, the motion of the charged
particles is Landau quantized in the direction perpendicular to the
magnetic field.  We choose the coordinate axes in such a way that ${B}$
is along $z$-axis. Then, the single particle energy of any charged particle at 
$n$-th Landau level is given by
\begin{equation}
\label{landau}
E_n =  \sqrt{p_z^2 + m^2 + 2 n e |Q|{B}},
\end{equation}
where $m$ is the mass of the particle, $p_z$ is the component of momentum
along $z$ direction.  The Landau levels in
Eq.~(\ref{landau}) assume integer values $n=0,1,2...$ for spin-up
states and $n=1,2,3...$ for spin-down states for positively
charged particles. For negatively charged particles $n$ takes 
on values $n=0,1,2,...$ for the spin-down states and $n=1,2,3...$ 
for spin-up states. The zero-temperature number density of 
charged baryons and leptons is given by
\begin{equation}
n_C=\frac {e|Q|{B}}{2\pi^2}\sum_{n=0}^{n_{max}}(2-\delta_{n,0})
\sqrt{p_F^2 -2ne|Q|B},
\end{equation}
with
\begin{equation}
n_{max}=\mathrm {Int}\left(\frac{p_F^2}{2e|Q|{B}}\right),
\end{equation}
where $p_F$ is the Fermi momentum. At zero temperature 
the number density of neutral baryons is expressed via their Fermi
momentum as 
\begin{equation}
n_N=\frac{p_F^3}{3\pi^2}.
\end{equation}
The chemical potentials of baryons are given by  $\mu_B=\mu_B^*
+\omega^0g_{\omega B}+\rho_3^0I_3^{(B)}g_{\rho B}$, where 
\begin{equation}
\mu_B^*~= ~\sqrt{p_F^{(B)^2}+m_B^{*^2}},
\end{equation}
with $m_B^*$ being the effective mass of the baryon $B$ in the mean
field approximation. We will further need the quantities 
\begin{eqnarray}
\omega^0 &=& \frac1{m_\omega^2}\sum_B g_{\omega B} n_B, \\
\rho_3^0 &=& \frac1{m_\rho^2}\sum_B g_{\rho B} I_3^{(B)} n_B,
\end{eqnarray}
which are the ground state expectation values of $\omega$ and 
$\rho_3$ fields with $I_3^{(B)}$ being the isospin projection for 
the baryon $B$. The chemical potentials of leptons are given by 
\begin{equation}
\mu_l~= ~\sqrt{{p_F^{(l)^2}}+m_l^2},
\end{equation}
where $p_F^{(l)}$ and $m_l$ are the Fermi momenta and masses of
leptons, respectively.  Now we are in the position to write down the
matter energy density defined by the matter Lagrangian density, ${\cal L}_m$, given
by Eq. (\ref{lag}) 
\begin{eqnarray}
\varepsilon_m &=&  \frac1{8 \pi^2} \sum_B 
\left(2 {p_F^{(B)}\mu_B^{*^3}}
 - p_F^{(B)}m_B^{*^2}\mu_B^* 
 - m_B^{*^4}~\ln \frac{p_F^{(B)}+\mu_B^*}{m^*_B} \right) \nonumber\\
&+&   \frac{e|Q|B}{(2\pi)^2}\sum_{B'}  \sum_{n=0}^{n_{max}} (2-\delta_{n,0})
\left[p_{B'}(n)\mu_{B'}^*~+~( m_{B'}^{*^2} + 2ne|Q|B) 
\ln \left(\frac{p_{B'}(n)+\mu_{B'}^*}{\sqrt {m_{B'}^{*^2} +
      2ne|Q|B}}\right)
\right] \nonumber\\
&+&  \frac{e|Q|B}{(2\pi)^2} \sum_{l=e,\mu} \sum_{n=0}^{n_{max}}
(2-\delta_{n,0}) \left[p_l(n)\mu_l~+~(m_l^2 + 2ne|Q|B) 
\ln \left(\frac{p_l(n)+\mu_l}{\sqrt{m_l^2 +
      2ne|Q|B}}\right)\right]\nonumber\\
&+&\frac12 m_\sigma^2 \sigma^2 + U(\sigma)
+ \frac12 m_\omega^2 \omega^{0^2}+\frac12 m_\rho^2 \rho_3^{0^2},
\label{enden}
\end{eqnarray}
where the sums over $B$ and $B'$ include the uncharged and charged
baryons, respectively, and the following short-hand notation has been used
\begin{equation}
p(n)~=~\sqrt{p_F^2-2ne|Q|B}. 
\end{equation}
The matter pressure can be easily constructed from Eq. (\ref{enden})
using the thermodynamic relation 
\begin{equation}
P=\sum_B \mu_B n_B + \sum_l \mu_l n_l - \varepsilon_m.
\label{press}
\end{equation} 
We now account for the fact that the matter in compact stars is charge
neutral and in beta equilibrium.  The first requirement relates the
partial densities of charged particles according to 
\begin{equation}
\label{charge_neutrality}
n_p + n_{\Sigma^+} = n_{\Sigma^-} + n_{\Xi^-} + n_e + n_\mu.
\end{equation}
Equilibrium with respect to weak interactions further implies the
following relations 
\begin{eqnarray}
\mu_p = \mu_n - \mu_e,\quad
\mu_\mu = \mu_e,\quad
\mu_\Lambda = \mu_n,\quad 
\mu_{\Sigma^-} = \mu_n + \mu_e,\nonumber\\
\label{chem_eq2}
\mu_{\Sigma^0} = \mu_n,\quad
\mu_{\Sigma^+} = \mu_p,\quad 
\mu_{\Xi^-} = \mu_n + \mu_e,\quad 
\mu_{\Xi^0} = \mu_n.
\label{chem_eq}
\end{eqnarray}
These conditions are normalized such that the total number of baryons
is reproduced, \ie 
\begin{equation}
\label{normalization}
n_b=\sum_B n_B.
\end{equation}
The solutions of the field equations at any given baryon density 
$n_B$  and zero
temperature are found in the mean-field approximation under the
constraints (\ref{charge_neutrality}), (\ref{chem_eq}), and
(\ref{normalization}).  Subsequently, the energy density and the
thermodynamic pressure of the matter are obtained using
Eqs. (\ref{enden}) and (\ref{press}).

In the rest frame of the hadronic fluid, with $B$ field along the $z$
axis, the matter and field parts of the energy-stress tensor are given,
respectively,  by 
\begin{equation}
\label{eq:Tmatter}
T^{\mu\nu}_m = \left(\begin{array}{cccc} 
                    \varepsilon_m & 0 & 0 & 0\\
                    0 & P-MB & 0 & 0\\
                    0 & 0 & P-MB & 0\\
                    0 & 0 & 0 & P 
               \end{array}\right),
\end{equation}
\begin{equation}
\label{eq:Tfield}
T^{\mu\nu}_f = \frac{B^2}{8\pi} \left(\begin{array}{cccc} 
                    1 & 0 & 0 & 0\\
                    0 & 1 & 0 & 0\\
                    0 & 0 & 1& 0\\
                    0 & 0 & 0 &-1 
               \end{array}\right).
\end{equation}
The total energy density of the system is given by the sum of the
matter and field contributions
\begin{equation}
\varepsilon = \varepsilon_m + \frac{B^2}{8\pi}.
\end{equation}
It is seen  from Eqs. ~(\ref{eq:Tmatter}) and ~(\ref{eq:Tfield}) that the 
the pressure in the perpendicular direction to the magnetic field is 
\begin{equation}
P_\perp = P - MB + \frac{B^2}{8\pi},
\label{perp}
\end{equation}
and the pressure in the direction parallel  to the magnetic field is given by
\begin{equation}
P_\parallel = P - \frac{B^2}{8\pi}.
\label{par}
\end{equation}
The couplings in the hypernuclear Lagrangian are fixed as follows.  In
the nucleonic sector the nucleon-meson coupling constants are chosen
according to Refs.~\cite{1987ZPhyA.327..295G,1987ZPhyA.326...57G,1982PhLB..114..392G,1985ApJ...293..470G} and
reproduce the bulk properties of nuclear matter (the binding energy
$E/B=-16.3$ MeV, the saturation density $n_0=0.153$ fm$^{-3}$, the
asymmetry energy coefficient $a_{asy}=32.5$ MeV, the incompressibility
$K=240$ MeV and the effective nucleon mass at the saturation $m^*/m =
0.8$).  In the hyperonic
sector, the hyperon-$\omega$ coupling constants is derived from the
$SU(6)$ symmetry of the quark model~\cite{Dover:1983yu,1994AnPhy.235...35S,1996PhRvC..53.1416S}.  To infer the
hyperon-$\sigma$ coupling constants we write the potential depth of a
hyperon $Y$ in nuclear matter at saturation as
\begin{equation}
U_Y=-g_{\sigma Y} \sigma+g_{\omega Y} \omega_0,
\end{equation}
and assign certain value to the potential $U_Y$, which then fixes the
value $g_{\sigma Y} $. We take the potential depth for
$\Lambda$-hyperon $U_\Lambda=-30~$MeV, as obtained from the analysis
of $\Lambda$-hypernuclei~\cite{Dover:1983yu,1998PhRvC..58.1306F}. 
For $\Xi$-hyperon we assign
$U_\Xi=-18$~MeV according to experimental data of
Refs.~\cite{1998PhRvC..58.1306F,2000PhRvC..61e4603K}. According to the recent $\Sigma$-hypernuclei
data the potential for $\Sigma$ hyperons is repulsive~\cite{1999PhRvL..83.5238B} 
and we adopt the value $U_\Sigma= 30$~MeV.

To model the density profile of the magnetic field we adopt the
following formula~\cite{1997PhRvL..79.2176B}
\begin{equation}
\label{profile}
B\left(\frac{n_b}{n_0}\right)=B_s+B_c\left\{1-\exp\left[{-\beta \left(
\frac {n_b}{n_0} \right)^\gamma}\right]\right\}.
\end{equation}
As discussed in the introduction the present formula allows us to
model the realistic physical situations where the magnetic field of
the star is non-uniform which is a realistic situation. The functional
dependence of the profile (\ref{profile}) on density is constructed
such as to take into account the fact that the magnetic field in the
cores of compact stars may be higher than that at the surface. The
parameters $\beta$ and $\gamma$ control the relaxation from the
central value $B_c$ to the asymptotic value at the suface
$B_s$. Obviously, the parameter $\beta$ controls the amount of field
decay at the saturation density, whereas the parameter $\gamma$
controls the width of the transition.  Observationally inferred value
of the surface magnetic field of magnetars is of the order of
$10^{15}$~G. It is plausible to asume that the central field could be in
the range of $10^{17}-10^{18}$~G.  Therefore, we will consider a
number of different values of $B_c$, while keeping $B_s =
10^{15}$~G. We have verified that the results are insensitive to the
precise value of the surface field $B_s$.

\section{Results and discussion}\label{discussion}

Our model of the equation of state of hypernuclear matter in strong
magnetic field is parameterized in terms of four physical quantities: the
boundary values of the field at the center and the surface, the amount
of its decline from the center to the saturation density (which is close
to the crust-core interface) and the steepness of the transition between the
two asymptotic values of the $B$-field. The latter two are described
by the parameters $\beta$ and $\gamma$, respectively. 
While the surface $B$-field has been fixed at the value of
$10^{15}$~G, we have explored a range of central magnetic fields; it
turns out that the equation of state is unaffected by the magnetic
fields below the value $B_c \le 10^{18}$~G.

\begin{figure*}
\begin{center}
\includegraphics[angle=0,width=12.0cm]{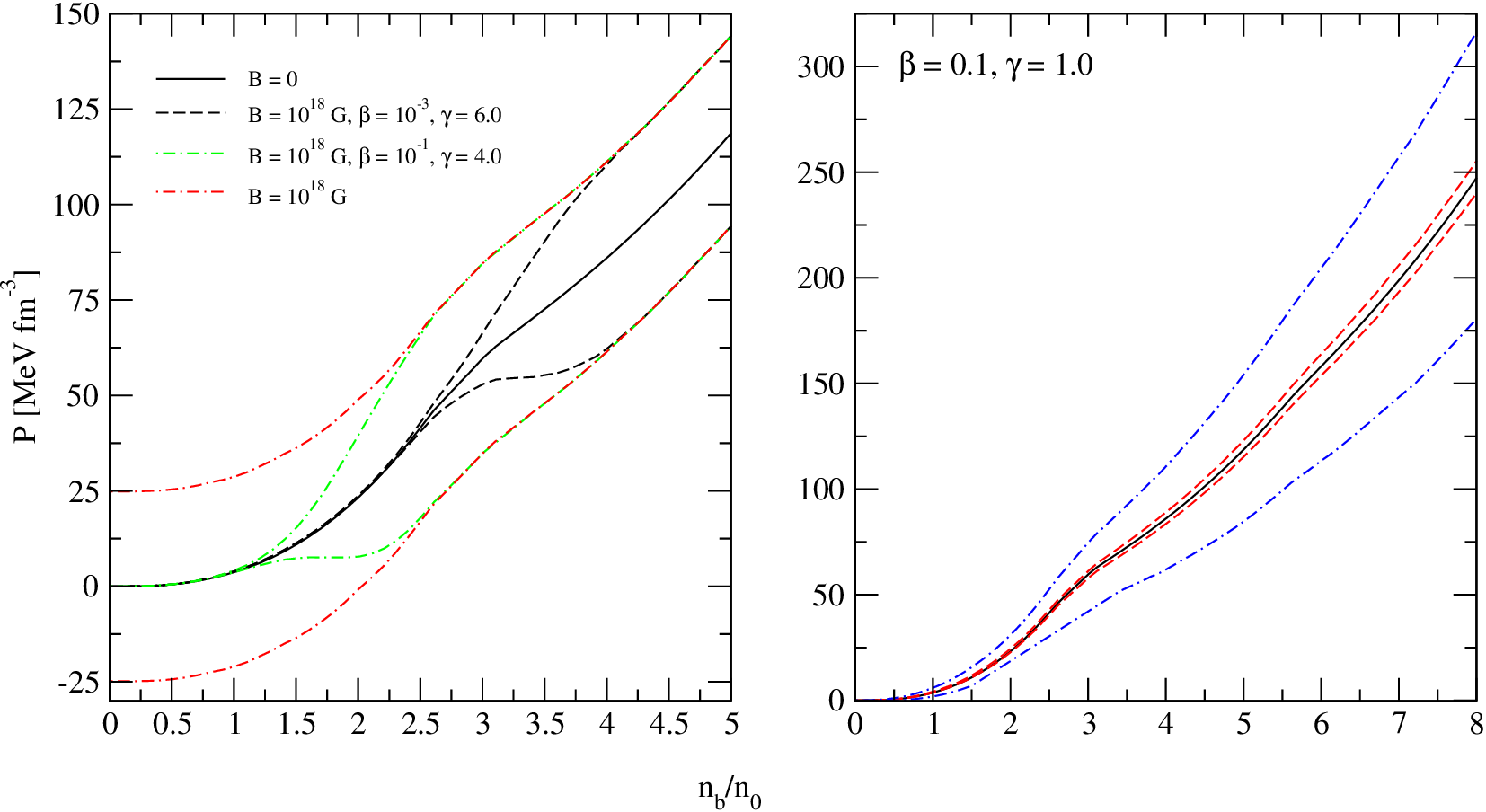}
\caption{ {\it Left panel:} Variation of total pressure as a function
  of normalized baryon number density for fixed magnetic fields $B_c =
  0$ and $B_c = 10^{18}$~G and several field profiles, $\beta =
  10^{-3}$, $\gamma = 6$ (dashed lines), $\beta = 10^{-1}$, $\gamma =
  4$ (dashed-double-dotted lines), and $\beta \to \infty$, \ie $B_c =$
  constant (dashed-dotted lines).  For each pair of curves the upper
  branch is for $P_\perp$ and the lower branch for $P_\parallel$.
  {\it Right panel:} Same as in the left panel, but for fixed $\beta =
  10^{-1}$, $\gamma = 1$ and for two values of $B$-field, $B_c = 10^{18}$
  G (dashed lines) and $B_c = 3\times 10^{18}$~G (dashed-dotted
  lines). The solid line corresponds to the case $B_c = 0$.  }
\label{eos1}
\end{center}
\end{figure*}

\begin{figure*}
\begin{center}
\includegraphics[angle=0,width=12.0cm]{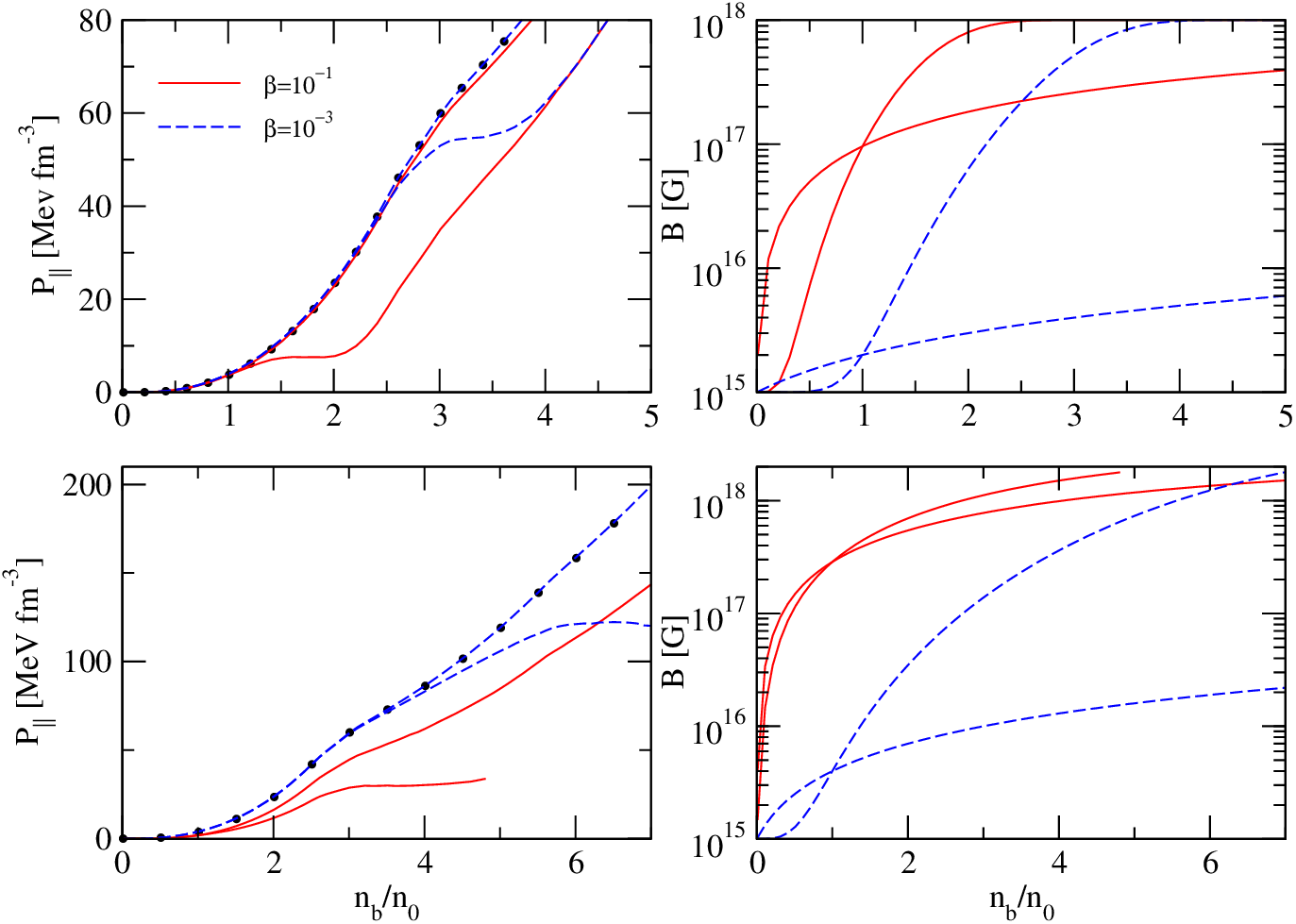}
\caption{ {\it Upper panels:} Dependence of $P_\parallel$ and $B$ on
  the normalized baryon number density for different magnetic field
  profiles and $B_c=10^{18}$ G. The dots show the reference case
  $B_c=0$.  The solid and dashed lines correspond to $\beta=0.1$ and
  $0.001$, respectively. For each $\beta$ we choose a pair of
  $\gamma$'s; in the first case we have $\gamma =1$ and $\gamma = 4$,
  whereas in the second case $\gamma = 1$ and $\gamma = 6$. The first
  $\gamma$ value for each $\beta$ corresponds to a less steeply rising
  curve.  {\it Lower panels:} Same as in the upper pannel but for
  $B_c=3\times10^{18}$~G and $\gamma = 1$ and $\gamma = 1.4$ for
  $\beta=0.1$ and $\gamma = 1$ and $\gamma = 3.5$ for
  $\beta=10^{-3}$. The specific values of $\gamma$'s were chosen to
  demonstrate the onset of instability.  }
\label{eos2}
\end{center}
\end{figure*}

Figure \ref{eos1} shows the equation of state of hypernuclear matter
in strong magnetic field.  The left panel shows the dependence of
pressure on the density for a fixed central $B$-field and various field 
profiles. For non-zero magnetic field the pressure splits into the
parallel and transverse components, which is evident from
Eqs. (\ref{perp}) and (\ref{par}), and is due to the gauge field
contribution.  The case with $\beta = 10^{-1}$ acquires a larger field
at the saturation density $n = n_0$, therefore the transition to the
asymptotic value of $B_c$ occurs earlier than in the case of $\beta =
10^{-3}$, both cases having approximately the same width of
transition (see Fig.~\ref{eos2}). It is clearly seen that 
the low-density behavior of the
equation of state with constant $B$-field implies unrealistically
large anisotropy of magnetic field up to the surface of the star,
which is inconsistent with the inferred surface magnetic field of
magnetars $B_s\sim 10^{15}$~G. The right panel of Fig.~\ref{eos1}
shows the equation of state in case of fixed density profile $\beta =
10^{-1}$ and $\gamma = 1$, but for various central magnetic fields. The
main effect seen in the figure is the splitting of the parallel and
transverse pressure as the field becomes larger, this effect being
sizable for fields above $10^{18}$~G.  Furthermore, we see that the
increase of $B_c$ causes the profile for $P_\parallel$ to become
softer;  the fact that $P_\parallel<P_\perp$ is due to negative
contribution of the field pressure to $P_\parallel$.

\begin{figure*}
\begin{center}
\includegraphics[angle=0,width=7.5cm]{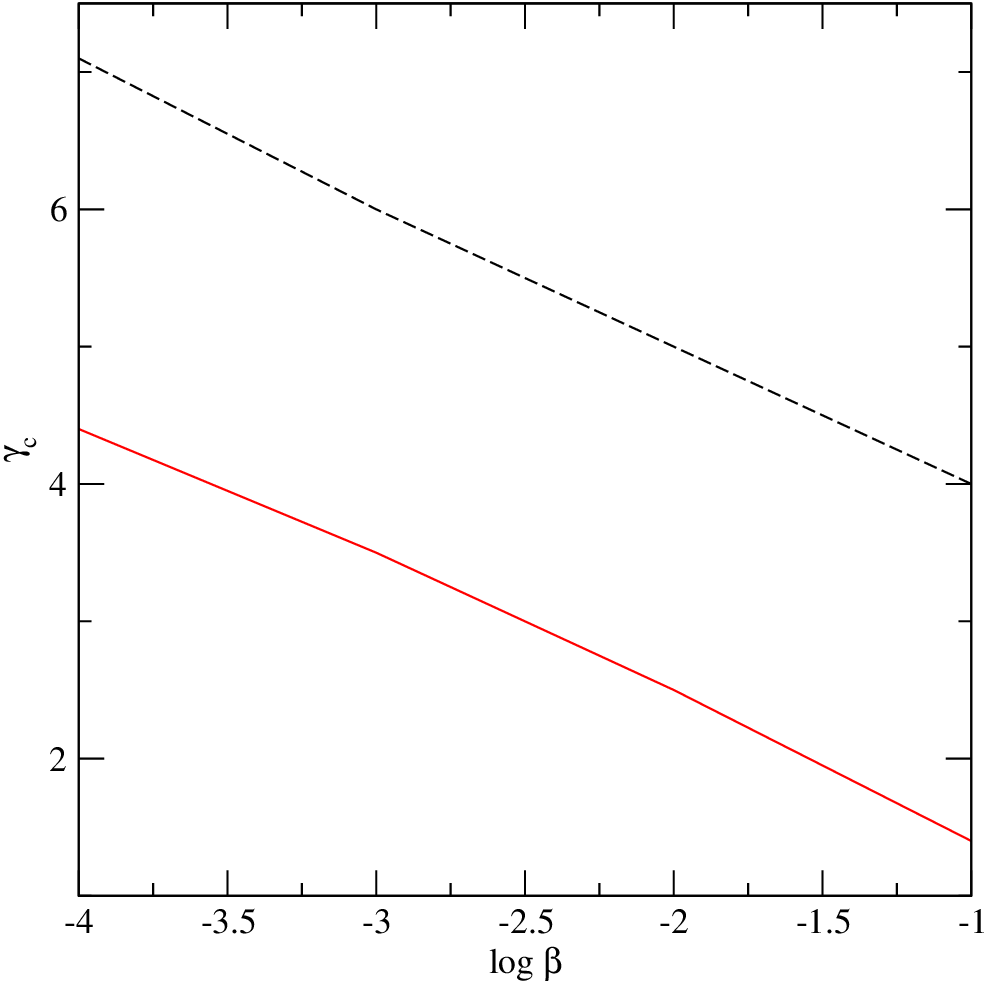}
\caption{Dependence of $\gamma_c$ - the critical value of $\gamma$
  parameter - on the logarithm of $\beta$ parameter. The solid line
  corresponds to the field $B_c = 3\times 10^{18}$~G, the dashed line - to
  $B_c = 10^{18}$~G }
\label{fig:gammac}
\end{center}
\end{figure*}

\begin{figure*}
\begin{center}
\includegraphics[angle=0,width=7.5cm]{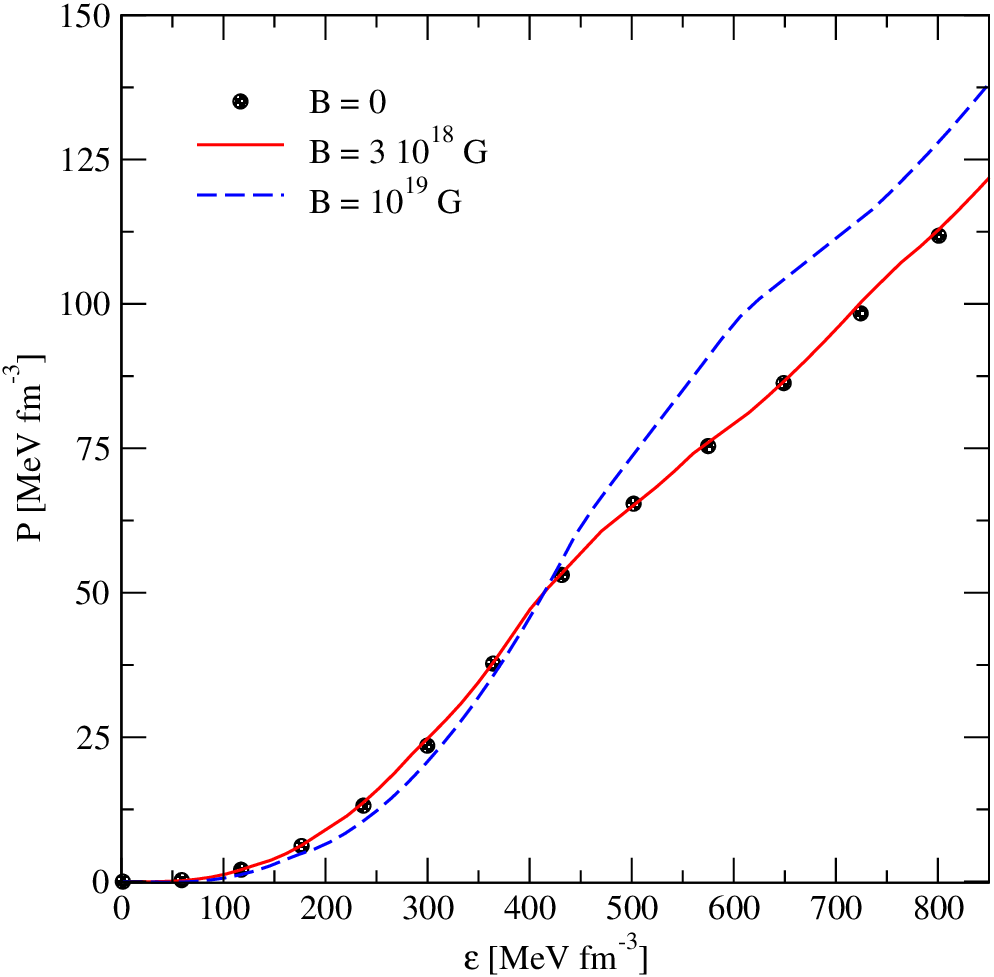}
\caption{Dependence of matter pressure on normalized 
baryon number density without the $B^2/8\pi$ 
term, which demonstrates the effect of Landau quantization.
}
\label{eos3}
\end{center}
\end{figure*}

\begin{figure*}
\begin{center}
\includegraphics[height=6.5cm,width=10.5cm]{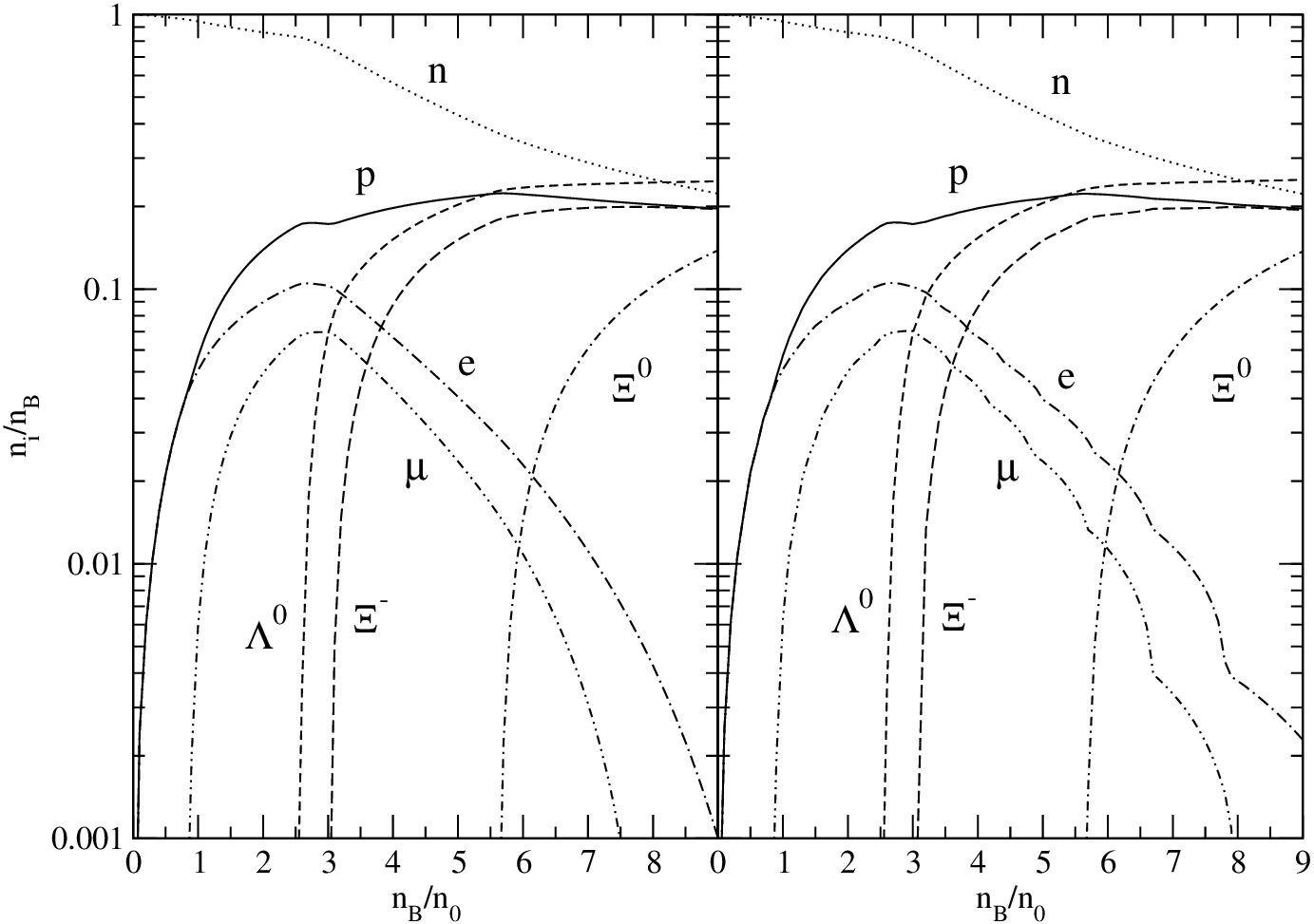}
\caption{ Abundances of different species as functions of normalized
  baryon number density without (left panel) and with (right panel)
  magnetic field $B=2\times 10^{18}$\, G.
The field profile parameters are $\beta =
  10^{-1}$, $\gamma = 1$, i.e., the field variations across the 
density profile of the star is as illustrated in Fig.~\ref{eos2}.
 }
\label{fig:abundances}
\end{center}
\end{figure*}

From the left panel of Fig.~\ref{eos1} we observe that there is an
onset of matter instability for chosen combinations of $\beta$ and
$\gamma$ (corresponding to different magnetic field profiles) for
fixed central field $B_c$.  This instability is induced in the
$P_\parallel$ component of the pressure, which decreases with the
density, because its field profile implies an increase of field strength.

To study the onset of instability in a more systematic manner, consider
the dependence of $P_\parallel$ on density for two central values of magnetic
field $B_c = 10^{18}$~G and $B_c = 3 \times 10^{18}$~G and several sets of
$\gamma$ and $\beta$ parameters. This is shown in the upper and lower
panels of Fig.~\ref{eos2}.  It is seen that for a given value of
$\beta$ the corresponding EoS becomes softer as the $\gamma$ is
increased.  Consequently, beyond a certain critical value of $\gamma$
and in a certain density regime $P_\parallel$ ceases to increase and
subsequently decreases with the further increase in $n_b$. This
implies that matter becomes unstable above that value of density for
that particular $B_c$ and magnetic field profile. This is evident from
Fig.~\ref{eos2}, where also the magnetic field profiles corresponding
to the onset of instability are shown.  For comparison we also show
results for each $\beta$ with the minimum value of $\gamma$ taken to be
$1$. Note that the maximum value of $\gamma$ is taken such that 
 $P_\parallel$ forms a plateau as a function $n_b$.
Furthermore, it is evident that with the decrease of $\beta$, the
instability occurs at larger values of $\gamma$ and $n_b$, which is
also evident from Fig~\ref{eos2}.

The instability arises due to the negative contribution from the field
energy density (pressure) to the pressures of magnetized baryons and
leptons in the direction to the magnetic field, which is evident from
Eq.~(\ref{par}). Since for any particular $B_c$ and magnetic profile,
the field strength increases with the increase of $n_b$, more negative
contribution is added to $P_\parallel$ with the increase of
$n_b$. Consequently, at a certain density, $P_\parallel$ ceases to
increase and then decreases with the increase of $n_b$. The critical
values of $\gamma$ for the onset of instability as a function of
log$\,\beta$ are shown in Fig.~\ref{fig:gammac}.

To demonstrate the effects of Landau quantization on the EoS we show
in Fig.~\ref{eos3} the EoS as a function of density with the field
energy density $B^2/8\pi$ subtracted. It is seen that the effects of
Landau quantization are insignificant even at field values $3\times
10^{18}$~G and become sizable only for fields of order
$10^{19}$~G.

Here we should mention that the true minimum which determines the
ground state of the matter is controlled by the Gibbs potential, which
contains an additional term $-\vecM\cdot\vecB$. However,
even at the strong field strength considered in the present study, the
magnetization effect is negligible. For example, the contribution of
the magnetization to $P_\perp$ in Eq. (\ref{perp}) is at least an
order of magnitude smaller than that due to matter and field,
depending upon the strength of the field and density. Hence, in our
numerical results, we do not consider the effect of magnetization.
However, it should be noted from Eqs. (\ref{perp}) and (\ref{par})
that, in principle, the magnetization leads to anisotropy of the total
pressure (apart from that due to the magnetic field itself) of the
system and contributes negatively to the pressure in the perpendicular
direction of the magnetic field. Therefore, inclusion of magnetization
further strengthens the main conclusion of this study.

In this context, it is also worth mentioning that the effect of
interaction of the anomalous magnetic moment (AMM) of particles with
field could stiffen EoS \cite{
2002PhLB..531..167B}. It could, in principle, dominate
over the softening effects by Landau quantization for fields strength
above $5\times10^{18}$ G \ie well above those considered in this
paper. It will start dominating over the Landau quantization effect
for fields $\sim 5\times 10^{19}$ G.  However, at such a high field
strength the field energy density contribution is more than enough to
overwhelm the AMM interaction.  Furthermore, the effect of AMM interaction
is density dependent~\cite{2002PhLB..531..167B}, and is most significant at low
densities.  In this respect, it is important to take a realistic
profile of a star, which will suppress the contribution from AMM
interactions because it implies low fields at low densities.

Finally, it is interesting to ask how the strong $B$-fields affect the
composition of matter. In Fig.~\ref{fig:abundances} we show the
composition of hypernuclear matter as functions of matter density in
cases with and without $B$-field. It is seen that the $B$-field has an
effect only on the lightest particles, \ie the leptons. For these we
see ``Landau oscillations'' in their population as a function of the
field. These oscillations reflect the population of new Landau levels
with the increase of the density. One should note that the field
varies across the density profile between the two asymptotic values at
the surface $10^{15}$ G and the center of the star $2\times 10^{18}$
G. The matter at low densities experiences relatively low field,
therefore the abundances in this region do not differ substantially
from the free field case. It is only in the high density region that
the field is large and appreciable deviations are seen for the leptons.

\section{Summary}
\label{summary}
We have studied the influence of strong magnetic fields on the high
density nuclear matter, including the possibile appearance of hyperons
at large densities.  In doing so we, first,  have {\it implemented realistic density
profiles of magnetic fields, which assume that the field decreases
from the center of the star and reaches asymptotically its value at
the surface $\sim 10^{15}$~G characteristic for magnetars}. Thus, we
have extended the previous studies of hypernuclear matter to realistic
situation of density dependent field profiles and conducted a
study of parameter space that defines the shape of the field profile.  

Secondly, we have found that for sufficiently large fields $B_c \ge
10^{18}$ G the matter becomes unstable.  {\it The instability is
  associated with the negative contribution of the field pressure to
  the baryonic and leptonic pressures in the direction parallel to the
  magnetic field,} which renders the total pressure of the system
anisotropic. We have found that the onset of instability depends on
the magnetic field profile (parameterized in terms of $\beta$ and
$\gamma$ parameters) as well as on the central field value $B_c$. The
instability depends on particular values and form of the density
profile, but it sets in always for critical central field values
$B_{\rm cr}\approx 10^{19}$ G for any values of $\beta$ and
$\gamma$. This gives a natural bound for the central magnetic field of
neutron stars with {\it homogeneously distributed magnetic field}.

\section*{Acknowledgement} M. S. acknowledges the support
 by the Alexander von Humboldt Foundation. B. M. acknowledges partial
 support through research Grant No. ISRO/RES/2/367/10-11.  We thank
 X.~G.~Huang and D. H. Rischke for discussions.


\begin{thebibliography}{10}
\expandafter\ifx\csname url\endcsname\relax
  \def\url#1{\texttt{#1}}\fi
\expandafter\ifx\csname urlprefix\endcsname\relax\def\urlprefix{URL }\fi
\expandafter\ifx\csname href\endcsname\relax
  \def\href#1#2{#2} \def\path#1{#1}\fi

\bibitem{1997ApJ...486L.129V}
G.~{Vasisht}, E.~V. {Gotthelf}, {The Discovery of an Anomalous X-Ray Pulsar in
  the Supernova Remnant Kes 73}, Astro. Phys. J. 486 (1997) L129.
\newblock \href {http://arxiv.org/abs/astro-ph/9706058}
  {\path{arXiv:astro-ph/9706058}}, \href {http://dx.doi.org/10.1086/310843}
  {\path{doi:10.1086/310843}}.

\bibitem{1998Natur.393..235K}
C.~{Kouveliotou}, S.~{Dieters}, T.~{Strohmayer}, J.~{van Paradijs}, G.~J.
  {Fishman}, C.~A. {Meegan}, K.~{Hurley}, J.~{Kommers}, I.~{Smith}, D.~{Frail},
  T.~{Murakami}, {An X-ray pulsar with a superstrong magnetic field in the soft
  {$\gamma$}-ray repeater SGR1806 - 20}, Nature 393 (1998) 235--237.
\newblock \href {http://dx.doi.org/10.1038/30410} {\path{doi:10.1038/30410}}.

\bibitem{1999ApJ...519L.139W}
P.~M. {Woods}, C.~{Kouveliotou}, J.~{van Paradijs}, K.~{Hurley}, R.~M.
  {Kippen}, M.~H. {Finger}, M.~S. {Briggs}, S.~{Dieters}, G.~J. {Fishman},
  {Discovery of a New Soft Gamma Repeater, SGR 1627-41}, Astro. Phys. J. 519
  (1999) L139--L142.
\newblock \href {http://arxiv.org/abs/astro-ph/9903267}
  {\path{arXiv:astro-ph/9903267}}, \href {http://dx.doi.org/10.1086/312124}
  {\path{doi:10.1086/312124}}.

\bibitem{1992ApJ...392L...9D}
R.~C. {Duncan}, C.~{Thompson}, {Formation of very strongly magnetized neutron
  stars - Implications for gamma-ray bursts}, Astro. Phys. J. 392 (1992)
  L9--L13.
\newblock \href {http://dx.doi.org/10.1086/186413} {\path{doi:10.1086/186413}}.

\bibitem{1992Natur.357..472U}
V.~V. {Usov}, {Millisecond pulsars with extremely strong magnetic fields as a
  cosmological source of gamma-ray bursts}, Nature 357 (1992) 472--474.
\newblock \href {http://dx.doi.org/10.1038/357472a0}
  {\path{doi:10.1038/357472a0}}.

\bibitem{1995MNRAS.275..255T}
C.~{Thompson}, R.~C. {Duncan}, {The soft gamma repeaters as very strongly
  magnetized neutron stars - I. Radiative mechanism for outbursts}, Mon. Not.
  Roy. Astron. Soc. 275 (1995) 255--300.

\bibitem{1996ApJ...473..322T}
C.~{Thompson}, R.~C. {Duncan}, {The Soft Gamma Repeaters as Very Strongly
  Magnetized Neutron Stars. II. Quiescent Neutrino, X-Ray, and Alfven Wave
  Emission}, Astro. Phys. J. 473 (1996) 322.
\newblock \href {http://dx.doi.org/10.1086/178147} {\path{doi:10.1086/178147}}.

\bibitem{1960AZh....37..193A}
V.~A. {Ambartsumyan}, G.~S. {Saakyan}, {The Degenerate Superdense Gas of
  Elementary Particles}, Astron. Zh. 37 (1960) 193.

\bibitem{1985ApJ...293..470G}
N.~K. {Glendenning}, {Neutron stars are giant hypernuclei?}, Astro. Phys. J.
  293 (1985) 470--493.
\newblock \href {http://dx.doi.org/10.1086/163253} {\path{doi:10.1086/163253}}.

\bibitem{1991PhRvL..67.2414G}
N.~K. {Glendenning}, S.~A. {Moszkowski}, {Reconciliation of neutron-star masses
  and binding of the Lambda in hypernuclei}, Physical Review Letters 67 (1991)
  2414--1417.
\newblock \href {http://dx.doi.org/10.1103/PhysRevLett.67.2414}
  {\path{doi:10.1103/PhysRevLett.67.2414}}.

\bibitem{1999paln.conf.....W}
F.~{Weber} (Ed.), {Pulsars as astrophysical laboratories for nuclear and
  particle physics}, 1999.

\bibitem{2007PrPNP..58..168S}
A.~{Sedrakian}, {The physics of dense hadronic matter and compact stars},
  Progress in Particle and Nuclear Physics 58 (2007) 168--246.
\newblock \href {http://arxiv.org/abs/nucl-th/0601086}
  {\path{arXiv:nucl-th/0601086}}, \href
  {http://dx.doi.org/10.1016/j.ppnp.2006.02.002}
  {\path{doi:10.1016/j.ppnp.2006.02.002}}.

\bibitem{1999ApJS..121..515B}
S.~{Balberg}, I.~{Lichtenstadt}, G.~B. {Cook}, {Roles of Hyperons in Neutron
  Stars}, Astro. Phys. J. 121 (1999) 515--531.
\newblock \href {http://arxiv.org/abs/astro-ph/9810361}
  {\path{arXiv:astro-ph/9810361}}, \href {http://dx.doi.org/10.1086/313196}
  {\path{doi:10.1086/313196}}.

\bibitem{2010PhRvC..81c5803D}
H.~{Djapo}, B.-J. {Schaefer}, J.~{Wambach}, {Appearance of hyperons in neutron
  stars}, Phys. Rev. C 81~(3) (2010) 035803.
\newblock \href {http://arxiv.org/abs/0811.2939} {\path{arXiv:0811.2939}},
  \href {http://dx.doi.org/10.1103/PhysRevC.81.035803}
  {\path{doi:10.1103/PhysRevC.81.035803}}.

\bibitem{2003astro.ph.12446B}
M.~{Baldo}, G.~F. {Burgio}, H.~. {Schulze}, {Neutron Star Structure with
  Hyperons and Quarks}, ArXiv Astrophysics e-prints\href
  {http://arxiv.org/abs/astro-ph/0312446} {\path{arXiv:astro-ph/0312446}}.

\bibitem{2011EL.....9411002V}
I.~{Vida{\~n}a}, D.~{Logoteta}, C.~{Provid{\^e}ncia}, A.~{Polls}, I.~{Bombaci},
  {Estimation of the effect of hyperonic three-body forces on the maximum mass
  of neutron stars}, EPL (Europhysics Letters) 94 (2011) 11002.
\newblock \href {http://arxiv.org/abs/1006.5660} {\path{arXiv:1006.5660}},
  \href {http://dx.doi.org/10.1209/0295-5075/94/11002}
  {\path{doi:10.1209/0295-5075/94/11002}}.

\bibitem{2012A&A...539A..16B}
L.~{Bonanno}, A.~{Sedrakian}, {Composition and stability of hybrid stars with
  hyperons and quark color-superconductivity}, Astron. and Astrophys. 539
  (2012) A16.
\newblock \href {http://arxiv.org/abs/1108.0559} {\path{arXiv:1108.0559}},
  \href {http://dx.doi.org/10.1051/0004-6361/201117832}
  {\path{doi:10.1051/0004-6361/201117832}}.

\bibitem{2012ApJ...756...56J}
W.-Z. {Jiang}, B.-A. {Li}, L.-W. {Chen}, {Large-mass Neutron Stars with
  Hyperonization}, Astrophys. J. 756 (2012) 56.
\newblock \href {http://arxiv.org/abs/1207.1686} {\path{arXiv:1207.1686}},
  \href {http://dx.doi.org/10.1088/0004-637X/756/1/56}
  {\path{doi:10.1088/0004-637X/756/1/56}}.

\bibitem{2012A&A...543A.157B}
I.~{Bednarek}, P.~{Haensel}, J.~L. {Zdunik}, M.~{Bejger}, R.~{Ma{\'n}ka},
  {Hyperons in neutron-star cores and a 2 M$_{sun;}$ pulsar}, Astron. and
  Astrophys. 543 (2012) A157.
\newblock \href {http://arxiv.org/abs/1111.6942} {\path{arXiv:1111.6942}},
  \href {http://dx.doi.org/10.1051/0004-6361/201118560}
  {\path{doi:10.1051/0004-6361/201118560}}.

\bibitem{2012PhRvC..85e5806O}
M.~{Oertel}, A.~F. {Fantina}, J.~{Novak}, {Extended equation of state for
  core-collapse simulations}, Phys. Rev. C 85~(5) (2012) 055806.
\newblock \href {http://arxiv.org/abs/1202.2679} {\path{arXiv:1202.2679}},
  \href {http://dx.doi.org/10.1103/PhysRevC.85.055806}
  {\path{doi:10.1103/PhysRevC.85.055806}}.

\bibitem{2012arXiv1205.3621M}
K.~{Masuda}, T.~{Hatsuda}, T.~{Takatsuka}, {Hadron-Quark Crossover and Massive
  Hybrid Stars with Strangeness}, ArXiv e-prints\href
  {http://arxiv.org/abs/1205.3621} {\path{arXiv:1205.3621}}.

\bibitem{2011arXiv1112.6430L}
R.~{Lastowiecki}, D.~{Blaschke}, H.~{Grigorian}, S.~{Typel}, {Strangeness in
  the cores of neutron stars}, ArXiv e-prints\href
  {http://arxiv.org/abs/1112.6430} {\path{arXiv:1112.6430}}.

\bibitem{2011arXiv1112.1853S}
S.~{Schramm}, R.~{Negreiros}, J.~{Steinheimer}, T.~{Sch{\"u}rhoff},
  V.~{Dexheimer}, {Properties and Stability of Hybrid Stars}, ArXiv
  e-prints\href {http://arxiv.org/abs/1112.1853} {\path{arXiv:1112.1853}}.

\bibitem{2012arXiv1206.3086D}
V.~{Dexheimer}, J.~{Steinheimer}, R.~{Negreiros}, S.~{Schramm}, {Hybrid Stars
  in an SU(3) Parity Doublet Model}, ArXiv e-prints\href
  {http://arxiv.org/abs/1206.3086} {\path{arXiv:1206.3086}}.

\bibitem{2012NuPhA.881...62W}
S.~{Weissenborn}, D.~{Chatterjee}, J.~{Schaffner-Bielich}, {Hyperons and
  massive neutron stars: The role of hyperon potentials}, Nuclear Physics A 881
  (2012) 62--77.
\newblock \href {http://arxiv.org/abs/1111.6049} {\path{arXiv:1111.6049}},
  \href {http://dx.doi.org/10.1016/j.nuclphysa.2012.02.012}
  {\path{doi:10.1016/j.nuclphysa.2012.02.012}}.

\bibitem{2012PhRvC..85f5802W}
S.~{Weissenborn}, D.~{Chatterjee}, J.~{Schaffner-Bielich}, {Hyperons and
  massive neutron stars: Vector repulsion and SU(3) symmetry}, Phys. Rev. C
  85~(6) (2012) 065802.
\newblock \href {http://arxiv.org/abs/1112.0234} {\path{arXiv:1112.0234}},
  \href {http://dx.doi.org/10.1103/PhysRevC.85.065802}
  {\path{doi:10.1103/PhysRevC.85.065802}}.

\bibitem{2012arXiv1209.0270G}
F.~{Gulminelli}, A.~R. {Raduta}, J.~{Margueron}, P.~{Papakonstantinou},
  M.~{Oertel}, {Phase diagram of neutron-rich nuclear matter and its impact on
  astrophysics}, ArXiv e-prints\href {http://arxiv.org/abs/1209.0270}
  {\path{arXiv:1209.0270}}.

\bibitem{2012arXiv1207.4872M}
R.~{Mallick}, {Maximum mass of a hybrid star having a mixed phase region in the
  light of pulsar PSR J1614-2230}, ArXiv e-prints\href
  {http://arxiv.org/abs/1207.4872} {\path{arXiv:1207.4872}}.

\bibitem{1977FCPh....2..203C}
V.~{Canuto}, J.~{Ventura}, {Quantizing Magnetic Fields in Astrophysics},
  Fundam. Cosmic Phys. 2 (1977) 203--353.

\bibitem{1989ApJ...342..958F}
I.~{Fushiki}, E.~H. {Gudmundsson}, C.~J. {Pethick}, {Surface structure of
  neutron stars with high magnetic fields}, Astro. Phys. J. 342 (1989)
  958--975.
\newblock \href {http://dx.doi.org/10.1086/167653} {\path{doi:10.1086/167653}}.

\bibitem{1991ApJ...374..652A}
A.~M. {Abrahams}, S.~L. {Shapiro}, {Equation of state in a strong magnetic
  field - Finite temperature and gradient corrections}, Astro. Phys. J. 374
  (1991) 652--667.
\newblock \href {http://dx.doi.org/10.1086/170151} {\path{doi:10.1086/170151}}.

\bibitem{1992AnPhy.216...29F}
I.~{Fushiki}, E.~H. {Gudmundsson}, C.~J. {Pethick}, J.~{Yngvason}, {Matter in a
  magnetic field in the Thomas-Fermi and related theories}, Annals of Physics
  216 (1992) 29--72.
\newblock \href {http://dx.doi.org/10.1016/0003-4916(52)90041-9}
  {\path{doi:10.1016/0003-4916(52)90041-9}}.

\bibitem{1993ApJ...416..276R}
O.~E. {Roegnvaldsson}, I.~{Fushiki}, E.~H. {Gudmundsson}, C.~J. {Pethick},
  J.~{Yngvason}, {Thomas-Fermi Calculations of Atoms and Matter in Magnetic
  Neutron Stars: Effects of Higher Landau Bands}, Astro. Phys. J. 416 (1993)
  276.
\newblock \href {http://dx.doi.org/10.1086/173234} {\path{doi:10.1086/173234}}.

\bibitem{1997PhRvL..78.2898C}
S.~{Chakrabarty}, D.~{Bandyopadhyay}, S.~{Pal}, {Dense Nuclear Matter in a
  Strong Magnetic Field}, Physical Review Letters 78 (1997) 2898--2901.
\newblock \href {http://arxiv.org/abs/astro-ph/9703034}
  {\path{arXiv:astro-ph/9703034}}, \href
  {http://dx.doi.org/10.1103/PhysRevLett.78.2898}
  {\path{doi:10.1103/PhysRevLett.78.2898}}.

\bibitem{1998PhRvD..58l1301B}
D.~{Bandyopadhyay}, S.~{Chakrabarty}, P.~{Dey}, S.~{Pal}, {Rapid cooling of
  magnetized neutron stars}, Phys. Rev. D 58~(12) (1998) 121301.
\newblock \href {http://arxiv.org/abs/astro-ph/9804145}
  {\path{arXiv:astro-ph/9804145}}, \href
  {http://dx.doi.org/10.1103/PhysRevD.58.121301}
  {\path{doi:10.1103/PhysRevD.58.121301}}.

\bibitem{2000ApJ...537..351B}
A.~{Broderick}, M.~{Prakash}, J.~M. {Lattimer}, {The Equation of State of
  Neutron Star Matter in Strong Magnetic Fields}, Astro. Phys. J. 537 (2000)
  351--367.
\newblock \href {http://arxiv.org/abs/astro-ph/0001537}
  {\path{arXiv:astro-ph/0001537}}, \href {http://dx.doi.org/10.1086/309010}
  {\path{doi:10.1086/309010}}.

\bibitem{2001ApJ...546.1126S}
I.-S. {Suh}, G.~J. {Mathews}, {Cold Ideal Equation of State for Strongly
  Magnetized Neutron Star Matter: Effects on Muon Production and Pion
  Condensation}, Astro. Phys. J. 546 (2001) 1126--1136.
\newblock \href {http://arxiv.org/abs/astro-ph/9912301}
  {\path{arXiv:astro-ph/9912301}}, \href {http://dx.doi.org/10.1086/318277}
  {\path{doi:10.1086/318277}}.

\bibitem{2006RPPh...69.2631H}
A.~K. {Harding}, D.~{Lai}, {Physics of strongly magnetized neutron stars},
  Reports on Progress in Physics 69 (2006) 2631--2708.
\newblock \href {http://arxiv.org/abs/astro-ph/0606674}
  {\path{arXiv:astro-ph/0606674}}, \href
  {http://dx.doi.org/10.1088/0034-4885/69/9/R03}
  {\path{doi:10.1088/0034-4885/69/9/R03}}.

\bibitem{2007MPLA...22..623C}
W.~{Chen}, P.-Q. {Zhang}, L.-G. {Liu}, {The Influence of the Magnetic Field on
  the Properties of Neutron Star Matter}, Modern Physics Letters A 22 (2007)
  623--629.
\newblock \href {http://arxiv.org/abs/astro-ph/0505113}
  {\path{arXiv:astro-ph/0505113}}, \href
  {http://dx.doi.org/10.1142/S0217732307023213}
  {\path{doi:10.1142/S0217732307023213}}.

\bibitem{2008JPhG...35l5201R}
A.~{Rabhi}, C.~{Provid{\^e}ncia}, J.~{Da Provid{\^e}ncia}, {Stellar matter with
  a strong magnetic field within density-dependent relativistic models},
  Journal of Physics G Nuclear Physics 35~(12) (2008) 125201.
\newblock \href {http://arxiv.org/abs/0810.3390} {\path{arXiv:0810.3390}},
  \href {http://dx.doi.org/10.1088/0954-3899/35/12/125201}
  {\path{doi:10.1088/0954-3899/35/12/125201}}.

\bibitem{1995A&A...301..757B}
M.~{Bocquet}, S.~{Bonazzola}, E.~{Gourgoulhon}, J.~{Novak}, {Rotating neutron
  star models with a magnetic field.}, Astron. Astrophysics 301 (1995) 757.
\newblock \href {http://arxiv.org/abs/gr-qc/9503044}
  {\path{arXiv:gr-qc/9503044}}.

\bibitem{2001ApJ...554..322C}
C.~Y. {Cardall}, M.~{Prakash}, J.~M. {Lattimer}, {Effects of Strong Magnetic
  Fields on Neutron Star Structure}, Astro. Phys. J. 554 (2001) 322--339.
\newblock \href {http://arxiv.org/abs/astro-ph/0011148}
  {\path{arXiv:astro-ph/0011148}}, \href {http://dx.doi.org/10.1086/321370}
  {\path{doi:10.1086/321370}}.

\bibitem{2012arXiv1207.4035F}
J.~{Frieben}, L.~{Rezzolla}, {Equilibrium models of relativistic stars with a
  toroidal magnetic field}, ArXiv e-prints\href
  {http://arxiv.org/abs/1207.4035} {\path{arXiv:1207.4035}}.

\bibitem{2012PhRvD..86d2001D}
U.~{Das}, B.~{Mukhopadhyay}, {Strongly magnetized cold degenerate electron gas:
  Mass-radius relation of the magnetized white dwarf}, Phys. Rev. D 86~(4)
  (2012) 042001.
\newblock \href {http://arxiv.org/abs/1204.1262} {\path{arXiv:1204.1262}},
  \href {http://dx.doi.org/10.1103/PhysRevD.86.042001}
  {\path{doi:10.1103/PhysRevD.86.042001}}.

\bibitem{2012IJMPD..2142001D}
U.~{Das}, B.~{Mukhopadhyay}, {Violation of Chandrasekhar Mass Limit: the
  Exciting Potential of Strongly Magnetized White Dwarfs}, International
  Journal of Modern Physics D 21 (2012) 42001.
\newblock \href {http://arxiv.org/abs/1205.3160} {\path{arXiv:1205.3160}},
  \href {http://dx.doi.org/10.1142/S0218271812420011}
  {\path{doi:10.1142/S0218271812420011}}.

\bibitem{2012MPLA...2750084K}
A.~{Kundu}, B.~{Mukhopadhyay}, {Mass of Highly Magnetized White Dwarfs
  Exceeding the Chandrasekhar Limit:. AN Analytical View}, Modern Physics
  Letters A 27 (2012) 50084.
\newblock \href {http://arxiv.org/abs/1204.1463} {\path{arXiv:1204.1463}},
  \href {http://dx.doi.org/10.1142/S0217732312500848}
  {\path{doi:10.1142/S0217732312500848}}.

\bibitem{2010ApJ...713.1073S}
R.~A. {Scalzo}, G.~{Aldering}, P.~{Antilogus}, S.~{Aragon}, C.~{Bailey},
  C.~{Baltay}, S.~{Bongard}, M.~{Buton}, C.~{Childress}, N.~{Chotard},
  Y.~{Copin}, H.~K. {Fakhouri}, A.~{Gal-Yam}, E.~{Gangler}, S.~{Hoyer},
  M.~{Kasliwal}, S.~{Loken}, P.~{Nugent}, R.~{Pain}, E.~{P{\'e}contal},
  R.~{Pereira}, S.~{Perlmutter}, D.~{Rabinowitz}, A.~{Rau}, G.~{Rigaudier},
  K.~{Runge}, G.~{Smadja}, C.~{Tao}, R.~C. {Thomas}, B.~{Weaver}, C.~{Wu},
  {Nearby Supernova Factory Observations of SN 2007if: First Total Mass
  Measurement of a Super-Chandrasekhar-Mass Progenitor}, Astro. Phys. J. 713
  (2010) 1073--1094.
\newblock \href {http://arxiv.org/abs/1003.2217} {\path{arXiv:1003.2217}},
  \href {http://dx.doi.org/10.1088/0004-637X/713/2/1073}
  {\path{doi:10.1088/0004-637X/713/2/1073}}.

\bibitem{1953ApJ...118..116C}
S.~{Chandrasekhar}, E.~{Fermi}, {Problems of Gravitational Stability in the
  Presence of a Magnetic Field.}, Astro. Phys. J. 118 (1953) 116.
\newblock \href {http://dx.doi.org/10.1086/145732} {\path{doi:10.1086/145732}}.

\bibitem{1991ApJ...383..745L}
D.~{Lai}, S.~L. {Shapiro}, {Cold equation of state in a strong magnetic field -
  Effects of inverse beta-decay}, Astro. Phys. J. 383 (1991) 745--751.
\newblock \href {http://dx.doi.org/10.1086/170831} {\path{doi:10.1086/170831}}.

\bibitem{2000PhRvL..84.5261C}
M.~{Chaichian}, S.~S. {Masood}, C.~{Montonen}, A.~{P{\'e}\ rez
  Mart{\'{\i}}nez}, H.~{P{\'e}rez Rojas}, {Quantum Magnetic Collapse}, Physical
  Review Letters 84 (2000) 5261--5264.
\newblock \href {http://arxiv.org/abs/hep-ph/9911218}
  {\path{arXiv:hep-ph/9911218}}, \href
  {http://dx.doi.org/10.1103/PhysRevLett.84.5261}
  {\path{doi:10.1103/PhysRevLett.84.5261}}.

\bibitem{2008IJMPD..17.2107P}
A.~{P{\'e}rez Mart{\'{\i}}nez}, H.~{P{\'e}rez Rojas}, H.~J. {Mosquera Cuesta},
  {Anisotropic Pressures in Very Dense Magnetized Matter}, International
  Journal of Modern Physics D 17 (2008) 2107--2123.
\newblock \href {http://arxiv.org/abs/0711.0975} {\path{arXiv:0711.0975}},
  \href {http://dx.doi.org/10.1142/S0218271808013741}
  {\path{doi:10.1142/S0218271808013741}}.

\bibitem{2010PhRvD..81d5015H}
X.-G. {Huang}, M.~{Huang}, D.~H. {Rischke}, A.~{Sedrakian}, {Anisotropic
  hydrodynamics, bulk viscosities, and r-modes of strange quark stars with
  strong magnetic fields}, Phys. Rev. D 81~(4) (2010) 045015.
\newblock \href {http://arxiv.org/abs/0910.3633} {\path{arXiv:0910.3633}},
  \href {http://dx.doi.org/10.1103/PhysRevD.81.045015}
  {\path{doi:10.1103/PhysRevD.81.045015}}.

\bibitem{2011AnPhy.326.3075H}
X.-G. {Huang}, A.~{Sedrakian}, D.~H. {Rischke}, {Kubo formulas for relativistic
  fluids in strong magnetic fields}, Annals of Physics 326 (2011) 3075--3094.
\newblock \href {http://arxiv.org/abs/1108.0602} {\path{arXiv:1108.0602}},
  \href {http://dx.doi.org/10.1016/j.aop.2011.08.001}
  {\path{doi:10.1016/j.aop.2011.08.001}}.

\bibitem{2010PhRvC..82f5802F}
E.~J. {Ferrer}, V.~{de La Incera}, J.~P. {Keith}, I.~{Portillo}, P.~L.
  {Springsteen}, {Equation of state of a dense and magnetized fermion system},
  Phys. Rev. C 82~(6) (2010) 065802.
\newblock \href {http://arxiv.org/abs/1009.3521} {\path{arXiv:1009.3521}},
  \href {http://dx.doi.org/10.1103/PhysRevC.82.065802}
  {\path{doi:10.1103/PhysRevC.82.065802}}.

\bibitem{2011PhRvD..83d3009P}
L.~{Paulucci}, E.~J. {Ferrer}, V.~{de La Incera}, J.~E. {Horvath}, {Equation of
  state for the magnetic-color-flavor-locked phase and its implications for
  compact star models}, Phys. Rev. D 83~(4) (2011) 043009.
\newblock \href {http://arxiv.org/abs/1010.3041} {\path{arXiv:1010.3041}},
  \href {http://dx.doi.org/10.1103/PhysRevD.83.043009}
  {\path{doi:10.1103/PhysRevD.83.043009}}.

\bibitem{2011arXiv1108.4479D}
V.~{Dexheimer}, R.~{Negreiros}, S.~{Schramm}, {Quark Deconfinement Under the
  Influence of Strong Magnetic Fields}, ArXiv e-prints\href
  {http://arxiv.org/abs/1108.4479} {\path{arXiv:1108.4479}}.

\bibitem{PhysRevC.85.039801}
A.~Y. Potekhin, D.~G. Yakovlev,
  \href{http://link.aps.org/doi/10.1103/PhysRevC.85.039801}{Comment on
  ``equation of state of a dense and magnetized fermion system''}, Phys. Rev. C
  85 (2012) 039801.
\newblock \href {http://dx.doi.org/10.1103/PhysRevC.85.039801}
  {\path{doi:10.1103/PhysRevC.85.039801}}.
\newline\urlprefix\url{http://link.aps.org/doi/10.1103/PhysRevC.85.039801}

\bibitem{2012PhRvC..85c9802F}
E.~J. {Ferrer}, V.~{de La Incera}, J.~P. {Keith}, I.~{Portillo}, P.~L.
  {Springsteen}, {Reply to ``Comment on `Equation of state of a dense and
  magnetized fermion system' ''}, Phys. Rev. C 85~(3) (2012) 039802.
\newblock \href {http://arxiv.org/abs/1110.0420} {\path{arXiv:1110.0420}},
  \href {http://dx.doi.org/10.1103/PhysRevC.85.039802}
  {\path{doi:10.1103/PhysRevC.85.039802}}.

\bibitem{1997PhRvL..79.2176B}
D.~{Bandyopadhyay}, S.~{Chakrabarty}, S.~{Pal}, {Quantizing Magnetic Field and
  Quark-Hadron Phase Transition in a Neutron Star}, Physical Review Letters 79
  (1997) 2176--2179.
\newblock \href {http://arxiv.org/abs/astro-ph/9703066}
  {\path{arXiv:astro-ph/9703066}}, \href
  {http://dx.doi.org/10.1103/PhysRevLett.79.2176}
  {\path{doi:10.1103/PhysRevLett.79.2176}}.

\bibitem{1974AnPhy..83..491W}
J.~D. {Walecka}, {A theory of highly condensed matter.}, Annals of Physics 83
  (1974) 491--529.
\newblock \href {http://dx.doi.org/10.1016/0003-4916(74)90208-5}
  {\path{doi:10.1016/0003-4916(74)90208-5}}.

\bibitem{1977NuPhA.292..413B}
J.~{Boguta}, A.~R. {Bodmer}, {Relativistic calculation of nuclear matter and
  the nuclear surface}, Nuclear Physics A 292 (1977) 413--428.
\newblock \href {http://dx.doi.org/10.1016/0375-9474(77)90626-1}
  {\path{doi:10.1016/0375-9474(77)90626-1}}.

\bibitem{1982PhLB..114..392G}
N.~K. {Glendenning}, {The hyperon composition of neutron stars}, Physics
  Letters B 114 (1982) 392--396.
\newblock \href {http://dx.doi.org/10.1016/0370-2693(82)90078-8}
  {\path{doi:10.1016/0370-2693(82)90078-8}}.

\bibitem{1987ZPhyA.326...57G}
N.~K. {Glendenning}, {Hyperons in neutron stars.}, Zeitschrift fur Physik A
  Hadrons and Nuclei 326 (1987) 57--64.

\bibitem{1987ZPhyA.327..295G}
N.~K. {Glendenning}, {Role of hyperons and pions in neutron stars and
  supernova}, Zeitschrift fur Physik A Hadrons and Nuclei 327 (1987) 295--300.
\newblock \href {http://dx.doi.org/10.1007/BF01284453}
  {\path{doi:10.1007/BF01284453}}.

\bibitem{1977AnPhy.108..301C}
S.~A. {Chin}, {A relativistic many-body theory of high density matter.}, Annals
  of Physics 108 (1977) 301--306.
\newblock \href {http://dx.doi.org/10.1016/0003-4916(77)90016-1}
  {\path{doi:10.1016/0003-4916(77)90016-1}}.

\bibitem{Serot:1984ey}
B.~D. Serot, J.~D. Walecka, {The Relativistic Nuclear Many Body Problem},
  Adv.Nucl.Phys. 16 (1986) 1--327.

\bibitem{2002PhRvD..65e6001K}
V.~R. {Khalilov}, {Macroscopic effects in cold magnetized nucleons and
  electrons with anomalous magnetic moments}, Phys. Rev. D 65~(5) (2002)
  056001.
\newblock \href {http://dx.doi.org/10.1103/PhysRevD.65.056001}
  {\path{doi:10.1103/PhysRevD.65.056001}}.

\bibitem{Dover:1983yu}
C.~Dover, A.~Gal, {SIGMA HYPERNUCLEI}, Comments Nucl.Part.Phys. 12 (1984)
  155--165.
\newblock \href {http://dx.doi.org/10.1016/0375-9474(85)90509-3}
  {\path{doi:10.1016/0375-9474(85)90509-3}}.

\bibitem{1994AnPhy.235...35S}
J.~{Schaffner}, C.~B. {Dover}, A.~{Gal}, C.~{Greiner}, D.~J. {Millener},
  H.~{Stocker}, {Multiply Strange Nuclear Systems}, Annals of Physics 235
  (1994) 35--76.
\newblock \href {http://dx.doi.org/10.1006/aphy.1994.1090}
  {\path{doi:10.1006/aphy.1994.1090}}.

\bibitem{1996PhRvC..53.1416S}
J.~{Schaffner}, I.~N. {Mishustin}, {Hyperon-rich matter in neutron stars},
  Phys. Rev. C 53 (1996) 1416--1429.
\newblock \href {http://arxiv.org/abs/nucl-th/9506011}
  {\path{arXiv:nucl-th/9506011}}, \href
  {http://dx.doi.org/10.1103/PhysRevC.53.1416}
  {\path{doi:10.1103/PhysRevC.53.1416}}.

\bibitem{1998PhRvC..58.1306F}
T.~{Fukuda}, A.~{Higashi}, Y.~{Matsuyama}, C.~{Nagoshi}, J.~{Nakano},
  M.~{Sekimoto}, P.~{Tlust{\'y}}, J.~K. {Ahn}, H.~{En'yo}, H.~{Funahashi},
  Y.~{Goto}, M.~{Iinuma}, K.~{Imai}, Y.~{Itow}, S.~{Makino}, A.~{Masaike},
  Y.~{Matsuda}, S.~{Mihara}, N.~{Saito}, R.~{Susukita}, S.~{Yokkaichi},
  K.~{Yoshida}, M.~{Yoshida}, S.~{Yamashita}, R.~{Takashima}, F.~{Takeutchi},
  S.~{Aoki}, M.~{Ieiri}, T.~{Iijima}, T.~{Yoshida}, I.~{Nomura}, T.~{Motoba},
  Y.~M. {Shin}, S.~{Weibe}, M.~S. {Chung}, I.~S. {Park}, K.~S. {Sim}, K.~S.
  {Chung}, J.~M. {Lee}, {Cascade hypernuclei in the (K$^{-}$,K$^{+}$) reaction
  on $^{12}$C}, Phys. Rev. C 58 (1998) 1306--1309.
\newblock \href {http://dx.doi.org/10.1103/PhysRevC.58.1306}
  {\path{doi:10.1103/PhysRevC.58.1306}}.

\bibitem{2000PhRvC..61e4603K}
P.~{Khaustov}, D.~E. {Alburger}, P.~D. {Barnes}, B.~{Bassalleck}, A.~R.
  {Berdoz}, A.~{Biglan}, T.~{B{\"u}rger}, D.~S. {Carman}, R.~E. {Chrien}, C.~A.
  {Davis}, H.~{Fischer}, G.~B. {Franklin}, J.~{Franz}, L.~{Gan}, A.~{Ichikawa},
  T.~{Iijima}, K.~{Imai}, Y.~{Kondo}, P.~{Koran}, M.~{Landry}, L.~{Lee},
  J.~{Lowe}, R.~{Magahiz}, M.~{May}, R.~{McCrady}, C.~A. {Meyer}, F.~{Merrill},
  T.~{Motoba}, S.~A. {Page}, K.~{Paschke}, P.~H. {Pile}, B.~{Quinn}, W.~D.
  {Ramsay}, A.~{Rusek}, R.~{Sawafta}, H.~{Schmitt}, R.~A. {Schumacher}, R.~W.
  {Stotzer}, R.~{Sutter}, F.~{Takeutchi}, W.~T. {van Oers}, K.~{Yamamoto},
  Y.~{Yamamoto}, M.~{Yosoi}, V.~J. {Zeps}, {Evidence of {$\Xi$} hypernuclear
  production in the $^{12}$C(K$^{-}$,K$^{+}$)$^{12}$$_{Ξ}$Be reaction}, Phys.
  Rev. C 61~(5) (2000) 054603.
\newblock \href {http://arxiv.org/abs/nucl-ex/9912007}
  {\path{arXiv:nucl-ex/9912007}}, \href
  {http://dx.doi.org/10.1103/PhysRevC.61.054603}
  {\path{doi:10.1103/PhysRevC.61.054603}}.

\bibitem{1999PhRvL..83.5238B}
S.~{Bart}, R.~E. {Chrien}, W.~A. {Franklin}, T.~{Fukuda}, R.~S. {Hayano},
  K.~{Hicks}, E.~V. {Hungerford}, R.~{Michael}, T.~{Miyachi}, T.~{Nagae},
  J.~{Nakano}, W.~{Naing}, K.~{Omata}, R.~{Sawafta}, Y.~{Shimizu}, L.~{Tang},
  S.~W. {Wissink}, {{$\Sigma$} Hyperons in the Nucleus}, Physical Review
  Letters 83 (1999) 5238--5241.
\newblock \href {http://dx.doi.org/10.1103/PhysRevLett.83.5238}
  {\path{doi:10.1103/PhysRevLett.83.5238}}.

\bibitem{2002PhLB..531..167B}
A.~E. {Broderick}, M.~{Prakash}, J.~M. {Lattimer}, {Effects of strong magnetic
  fields in strange baryonic matter}, Physics Letters B 531 (2002) 167--174.
\newblock \href {http://arxiv.org/abs/astro-ph/0111516}
  {\path{arXiv:astro-ph/0111516}}, \href
  {http://dx.doi.org/10.1016/S0370-2693(01)01514-3}
  {\path{doi:10.1016/S0370-2693(01)01514-3}}.

\end{thebibliography}

\end{document}